\documentclass{revtex4}
\def\[{\left\lbrack}
\def\]{\right\rbrack}

\def\({\left(}
\def\){\right)}
\newcommand{\be}{\begin{equation}}
\newcommand{\ee}{\end{equation}}
\newcommand{\ea}{\end{eqnarray}}
\newcommand{\ba}{\begin{eqnarray}}

\newcommand{\vx}{{\vec{x}}}
\newcommand{\vy}{{\vec{y}}}

\newcommand{\ep}{{\epsilon}}

\newcommand{\dirac}{{\delta(\vx - \vy)}}
\newcommand{\lnm}{{\cal L}_{\text{\tiny \it NCMCS}}}
\newcommand{\lns}{{\cal L}_{\text{\tiny \it NCSD}}}
\newcommand{\hfu}{\hat F^{\mu \nu}}
\newcommand{\hfd}{\hat F_{\mu \nu}}
\newcommand{\tf}{\tilde F}
\newcommand{\tfmu}{\tilde F^\mu}
\newcommand{\tfmd}{\tilde F_\mu}

\newcommand{\tth}{{\tilde \theta}}
\newcommand{\fmnu}{F^{\mu \nu}}
\newcommand{\fmnd}{F_{\mu \nu}}
\newcommand{\hamu}{\hat A^\mu}
\newcommand{\hamd}{\hat A_\mu}

\newcommand{\hand}{\hat A_\nu}

\newcommand{\hald}{\hat A_\lambda}
\newcommand{\amu}{A^\mu}
\newcommand{\amd}{A_\mu}

\newcommand{\ald}{A_\lambda}
\newcommand{\epu}{{\epsilon^{\mu \nu \lambda}}}
\newcommand{\epd}{{\epsilon_{\mu \nu \lambda}}}
\newcommand{\prt}{{\partial}}
\newcommand{\diag}{\mbox{diag}}

\begin{document}

\title{On duality of the noncommutative extension of the Maxwell-Chern-Simons model}

\author{M. S. Guimar\~aes, D. C. Rodrigues, C. Wotzasek}
\email{marcelosg, cabral, clovis@if.ufrj.br}
\affiliation{Instituto de F\'{\i}sica, Universidade Federal do Rio de Janeiro,\\
21945-970, Rio de Janeiro, RJ, Brasil}
\author{and J. L. Noronha}
\email{J.Noronha@figss.uni-frankfurt.de}
\affiliation{Frankfurt Institute for Advanced Studies (FIAS),
J.W.Goethe-Uni\\
Robert-Mayer-Str. 8-10
D-60054 Frankfurt am Main, Germany	}

\begin{abstract}
\vspace{0.7in}
\begin{center}
{\bf Abstract}
\end{center}
\noindent
We study issues of duality in 3D field theory models over a canonical noncommutative spacetime and obtain the noncommutative extension of the Self-Dual model induced by the Seiberg-Witten map.
We apply the dual projection technique to uncover some properties of the noncommutative Maxwell-Chern-Simons theory up to first-order in the noncommutative parameter. A duality between this theory and a model similar to the ordinary self-dual model is estabilished. The correspondence of the basic fields is obtained and the equivalence of algebras and equations of motion are directly verified. We also comment on previous results in this subject.
\end{abstract}

\maketitle


\setlength{\baselineskip} {20 pt}

\vspace{0.7in}
\section{Introduction}
This paper is devoted to study duality mapping and model equivalence in the context of three dimensional field theories over a canonical noncommutative spacetime (NC) \cite{rev}.
It is of great theoretical interest to speculate that the physical world might involve noncommutative coordinates and to ask about possible modifications to established concepts in the ordinary quantum field theory.
In particular the issue of duality is such a conspicuous notion in quantum field theory that it becomes mandatory to check if its consequences remain valid when considering NC-extensions of physically motivated theories and interesting models. Such studies have been indeed undergone for the NC-extensions of the electromagnetic 4D Maxwell theories and for the 3D NC-extension of the well known duality between Maxwell-Chern-Simons versus Self-Dual models.
Although for the last case we have seen a spate of studies in the recent literature, the results do not seem to agree which has motivated us to re-examine this issue.

By employing the Seiberg-Witten map (SWM)\cite{sw} we search for the dual companion of the NC-extension of the Maxwell-Chern-Simons model (NC-MCS) up to first-order in the non-commutative parameter $\theta$.
The results are therefore valid perturbatively for spacetimes with small noncommutativity.
This seems necessary since a map analog to the SWM is nonexistent for the Self-Dual model.
Therefore the basic strategy in this case has been to look for the NC-extension induced by the SWM over the MCS model.
To find the NC-extension of the Self-Dual model (NC-SD), i.e., the dual related model to the NC-MCS we employ the dual projection technique \cite{dual}.  Following this path we found the apropriate dual Lagrangians. Moreover the correspondence for the algebras of the observables and equations of motion were directly verified, expliciting the relation between fields of different models. It is worth recalling that the non-triviallity of the model equivalence studied here comes from the non-linear interactions of derivative type in the action due to the noncommutativity of the spacetime.

In ordinary space-time, the physical equivalence of MCS (topologically massive) \cite{mcs}, shown to represent a free massive spin one excitation and self-dual \cite{sd} theories has been proven quite useful by Deser and Jackiw in a seminal work \cite{dj}. In that paper, duality was first verified at the level of symplectic structures for MCS and SD models and then corroborated by use of the master action.  This duality equivalence seems important since the SD model was shown to appear in the bosonization of the fermionic massive Thirring model in the large wavelength limit \cite{fradschapo}. The Wilson loop operator of the dual gauge theory has a natural expression in terms of the fermion theory showing that a fermion loop operator may exhibit fractional statistics.
Planar gauge theories having excitations with arbitrary spin and statistics have also played important roles in the context of other physically interesting phenomena such as quantum Hall effect and high-TC superconductivity.

Recently several papers dealing with the extension of this duality to the noncommutative space have appeared \cite{dayi}\cite{cm}\cite{rob}\cite{ghosh} and the results found are quite distinct. 
The distinctiness seems to have its origin in the different techniques employed. For instance some authors
use the master (or interpolating) action approach. However in one case the master action is built for the commutative but nonlinear model after the SWM \cite{dayi} while in another instance the master action is obtained before the Seiberg-Witten map \cite{cm} running into the risk that an extension of the SWM might spoil the duality mapping. In both instances no check was done to see if the resulting actions provided the same set of field equations and/or physical observables. In \cite{rob} the duality for the NC-MCS was studied without employing the SWM. As so, the result is nonperturbative in $\theta$ and, consequently, it is difficult to directly compare that result with the basically perturbative approach of the other works.

A recent contribution \cite{ghosh} claims to have found duality as an example of a noncommutative free field theory in $(2+1)$- dimensions -- the abelian NC-MCS theory. In \cite{ghosh} by exploiting the Seiberg-Witten map, this result was argued to be expected since under the above mapping, the NC-Chern-Simons theory reduces to comutative Chern-Simons theory to all orders of $\theta$ and hence the results corresponding to commutative Chern-Simons theory should hold. 
It was also pointed out in \cite{ghosh} that no discussions on the symplectic structure of the theory or an explicit mapping between the degrees of freedom of the two purported dual theories have been attempted so far. It is true that by itself, relating the actions can not conclusively prove duality.
We agree with the criticism stated in \cite{ghosh} that of the use of a master Lagrangian to prove duality is not sufficient; for, although it can be a useful guide, a direct check is essential to assure the existence of duality.
The dual related actions obtained by any means should also go through some sort of confirmatory test of duality concerning the basic observables of theories. In fact the approach followed in \cite{ghosh} is very interesting -- after performing the SWM in the NC-MCS, the author derived the algebra of the observables for the resulting commutative but non-linear theory. By performing a ``sort of integration" of the algebra derived before the author was able to find an alternative representation of the algebra in terms of new vector fields.  He {\it correctly} interpreted that this new theory should be dual to the original NC-MCS. Therefore, based on established jargon, it should be named as the non-commutativy extension of the self-dual model. However, just like the works precceeding it, no attempt was done to check if the equations of motion coming from these independent representations of the common algebra indeed agree with one another.

In this work we employ another technique, known as dual projection \cite{dual}, to establish the correspondence between actions and fields, along with a direct verification of the equivalence of algebras and equations of motion.
The dual projection technique is a canonical transformation aiming to separate the field-variables responsible for the dynamical character of a given theory from those field-variables carrying the representation of the underlying symmetry. Consequently, this algorithm is able to provide not only the dual pair of actions but also the correspondence between the fields and, most importantly, to disclose the common algebra for the observable carried by both representations.

In this paper we adopt the dual projection programe to find the dual companion for the NC-MCS in $O(\theta)$ of the SWM and check for the consistency of the equations of motion and algebra for both representations. The next section is devoted to review the noncommutative MCS theory and the application of the Seiberg-Witten map. The resulting Lagrangian after the Seiberg-Witten map is analyzed under the dual projection approach and a new extension of the self-dual model to the noncomutative space is found. Afterwards,  the correspondence of algebras and equations of motion are verified. The last Section is reserved to the analysis of the results found and to our final remarks.

\vspace{0.4in}

\section{Duality in the Noncommutative Maxwell-Chern-Simons Theory}

The research in field theories based on spacetimes with intrinsic noncommutative coordinates \cite{snyder} has experienced a recent revival after the realization that this concept has a natural realization in string theory \cite{strings}.
In this framework, the commutator of the coordinates $x^\mu$ in the spacetime manifold is given by
\be
\label{theta}
[x^\mu, x^\nu] = i\,\theta^{\mu\nu}
\ee
where $\theta^{\mu\nu}$ is a constant real and antisymmetric matrix with dimensions of $(\mbox{length})^2$. 
One way to construct a noncommutative quantum field theory is to promote an established ordinary theory to a noncommutative one by replacing ordinary fields with noncommutative fields and ordinary products with Moyal *-products.
In the case of the Maxwell-Chern-Simons theory the noncommutative Lagrangian density is defined as \cite{dayi}\cite{gs} 
\be\label{ncmcs}
	\hat \lnm = - \frac 1 {4g} \hfu * \hfd + \frac m {2g} \epu \( \hamd * \prt_\nu \hald - \frac {2i}3 \hamd * \hand * \hald \),
\ee
with
\be
	\hfd \equiv \prt_\mu \hand - \prt_\nu \hamd - i \hamd * \hand + i \hand * \hamd,
\ee
$\mu, \nu, \lambda = 0,1,2$ and metric $(g_{\mu \nu}) = \diag \pmatrix { + & - & -}$. Here the constant $g$, with mass dimensions, is necessary in order to give dimensional consistency to the action, thus the field potentials $\hamu$ have dimension of mass and the Seiberg-Witten map can be applied without dimensional difficulties. The hat on a field means that its associated multiplication is not the ordinary one, but the *-product (ie, Moyal product), namely
\be
	(\hamd * \hand)(\vx) \equiv e^{\frac i2 \theta^{\alpha \beta} \prt^x_\alpha \prt^y_\beta}  \hamd(\vx) \hand(\vy)|_{\vy \rightarrow \vx} = \hamd(\vx) \hand(\vx) + \frac i2 \theta^{\alpha \beta} \prt_\alpha\hamd(\vx) \prt_\beta\hand(\vx) + O(\theta^2)\, ,
\ee
with $\theta^{\alpha\beta}$ defined as in (\ref{theta}).

The action of theory (\ref{ncmcs}) is invariant under the following infinitesimal gauge transformations
\be
	\hat \delta_{\hat \lambda} \hamu = \prt^\mu \hat \lambda + i \hat \lambda * \hamu - i  \hamu * \hat \lambda.
\ee
It is important to notice the factor $-2i/3$ that appears in the NC Chern-Simons term has a crucial role in regard to gauge symmetry; for the variation of the NC-Chern-Simons Lagrangian must be proportional to the field strength $\hfu$ \cite{gs}, otherwise gauge symmetry is lost and  there is a change in the number of degrees of freedom.

In what follows we shall resort to the Seiberg-Witten map, i.e., a correspondence between a noncommutative gauge theory and a conventional gauge theory to obtain, up to first-order in $\theta$, a commutative version of the theory (\ref{ncmcs}). The reason is that although it is in principle presumably possible to compute physical observables via noncommutative fields, the procedure leading to (\ref{ncmcs}) lacks direct information on how to identify realistic physical variables with specific operators.
This map connecting a noncommutative gauge theory with its commutative equivalent was proposed while analyzing open string theory in a magnetic field with two different regularization schemes.
It permits the construction of a commutative theory with ordinary gauge transformations having its physical content equivalent to the noncommutative theory.
Since then this notion has found many startling applications and connections to different branches of physics and mathematics.
The SWM ensures the stability of gauge transformations in the commutative and NC descriptions --  to ensure that a gauge transformation of $A^\mu$ is mapped to a noncommutative gauge transformation of $\hamu$, it becomes necessary that
\be
	\hamu(A) + \hat \delta_{\hat \lambda} \hamu (A) = \hamu (A + \delta_\lambda A)\, ,
\ee
whose solution, to first order in the noncommutative parameter $\theta$, leads to the map
\be
	\hamu = \amu - \theta^{\alpha \nu}  A_\alpha \(\prt_\nu \amu - \frac 12 \prt^\mu A_\nu \),
\ee
which implies
\be
	\hfd = F_{\mu \nu} + \theta^{\alpha \beta}(F_{\mu \alpha}F_{\nu \beta} - A_\alpha \prt_\beta F_{\mu \nu}).
\ee
The application of the map to the action given in (\ref{ncmcs}) results,  to first order in $\theta$,
\be
	g\lnm = -\frac 14 \fmnu \fmnd + 2 \theta^{\alpha \beta}(F_{\mu \alpha} F_{\nu \beta} \fmnu - \frac 14 F_{\alpha \beta} \fmnu \fmnd) + \frac m2 \epu \amd \prt_\nu \ald\, ,
\ee
which we still call as NC-MCS model as long as no risk of confusion with (\ref{ncmcs}) appears.

It is often claimed that noncommutative theories with $\theta^{0i}\neq 0$ may exhibit difficulties with perturbative unitarity while those ones with only $\theta^{ij}$ nonzero are acceptable \cite{uni,cas}. In odd dimensional spacetimes a totally antisymmetric matrix is necessarily singular therefore, due to Darboux theorem, it is always possible to find a coordinate system where at least one of the coordinates is a commuting one \cite{jfluid}\cite{poly}. We let this coordinate be associated with the time index, hence $\theta^{0i} = 0$.
Restricting ourselves to the commuting time case, the noncommutative extension of the MCS model, in first-order of $\theta$, gives \cite{ghosh}
\ba
	g\lnm &=& -\frac 14 \fmnu \fmnd + \frac m2 \epu \amd \prt_\nu \ald - \frac 18 \theta^{\alpha \beta} F_{\alpha \beta} \fmnu \fmnd \nonumber \\[0.2in]
	\label{e1}
	&=& - \frac 12 \( 1+ \theta \tilde F_0 \) \tilde F^\mu \tilde F_\mu + \frac m2 A^\mu \tilde F_\mu,
\ea
where $\theta \equiv \theta^{12}$ and $\tfmu \equiv \frac 12 \epu F_{\nu \lambda} = \epu \prt_\nu A_\lambda$.

Next we start to discuss the duality mapping. In order obtain the noncommutative extension of the self-dual model (NC-SD) we proceed with the dual projection \cite{dual} algorithm. To this end we introduce an auxiliary field $\pi^\mu$ as follows
\be
	\label{e2}
	g \lnm = \pi^\mu \tfmd + \frac 12 \( 1 -  \theta \tilde F_0 \) \pi^\mu \pi_\mu + \frac m2 A_\mu \tfmu
\ee
therefore lowering the order of the differential equations. The above procedure is just an ordinary Legendre transform, and the equivalence between (\ref{e1}) and (\ref{e2}) is easily verified by the substitution of the equations of motion of $\pi^\mu$ into (\ref{e2}),
\be
\label{dp1}
\pi_\mu  = - \( 1 +  \theta \tilde F_0 \) \tfmd
\ee
Next we disclose a canonical transformation aiming to diagonalize the action in such a way that one sector would be a pure gauge, carrying no propagating degrees of freedom. The other sector, carrying a representation of the dynamics, is therefore the interesting one for considerations of duality.  This will be done in two steps.  Firstly let us call
\be
	\label{dp2}
	\chi_\mu \equiv \pi_\mu - \frac 12 \theta \delta^0_\mu \pi^\alpha \pi_\alpha,
\ee
and then solve for $\pi = \pi (\chi)$ up to first-order in $\theta$ to eliminate the auxiliary field $\pi_\mu$ in favor of the new field-variable. Then
\be
	\label{e4}
	g \lnm = \( \chi^\mu + \frac m2 A^\mu \) \tfmd + \frac 12 ( 1 +  \theta \chi_0 ) \chi^\mu \chi_\mu.
\ee
Next, we define $p^\mu$ as a shift of $\chi^\mu$, namely
\be\label{dp3}
	p^\mu \equiv \chi^\mu + \frac m2 A^\mu,
\ee
in order to put the symplectic sector in a canonical form. Hence we can write
\be
	g \lnm = p^\mu \tfmd + \frac 12 \( p - \frac m2 A \)^\nu \( p - \frac m2 A \)_\nu \[1 + \theta \(p_0 - \frac m2 A_0 \) \].
\ee
We are now ready to complete the last step of the dual projection with the following canonical transformation
\ba\label{dp4}
	A_\mu &=& A^+_\mu + A^-_\mu, \nonumber\\
	p_\mu &=& \frac m2 (A^+_\mu - A^-_\mu),
\ea
that decouple the fields and diagonalizes the Lagrangian $\lnm$. The result of such redefinition is
\be
	g \lnm = \[\frac {m^2}2
(1 - m\theta A^-_0)A^-_\mu A^{-\mu} - \frac m2 \epu  A^-_\mu \partial_\nu A^-_\lambda\]
+  \[\frac m2 \epu A^+_\mu \partial_\nu A^+_\lambda \].
\ee
This factorization of the NC-MCS action into a pure Chern-Simons action for the $A^+$ fields and a dynamical action for the $A^-$ fields is an outstanding result.
The pure Chern-Simons term is surplus, it has no dynamical consequence and carries no propagating degrees of freedom.  It is responsible however for the gauge symmetry observed in the original model. The other part, the one with $A^-$ field is not a gauge theory. It carries the same dynamical content of the original NC-MCS being therefore dual to it. As so we name it as the noncommutative self-dual model, which reads
\be\label{ncsd}
	\lns =  \frac 1{2g} \( 1 -   \theta f_0 \) f_\mu f^{\mu} - \frac 1{2mg} \epu  f_\mu \partial_\nu f_\lambda,
\ee
after the replacement 
\be\label{dp5}
m A^-_\mu \rightarrow f_\mu \, .
\ee
It is interesting to observe that it correctly limits to the ordinary Self-Dual model when $\theta\to 0$.
This concludes the search for the noncommutative version of the Self-Dual model. As a bonus we may obtain directly from the dual projection procedure the correspondence among the basic field-variables of both models by tracing back the redefinitions done previously, Eqs.(\ref{dp1},\ref{dp2},\ref{dp3},\ref{dp4} and \ref{dp5}). The answer is 
\be
	\label{e10}
	f^\mu = \tf^\mu +  \tf^\mu \tf^\alpha \tth_\alpha + \frac 12 \tth^\mu \tf^\alpha  \tf_\alpha
\ee
and, therefore,
\be
	\label{e11}
	\tf^\mu = f^\mu -  f^\mu f^\alpha \tth_\alpha -\frac 12 \tth^\mu f^\alpha f_\alpha.
\ee
where  $\tth_\mu \equiv \frac 12 \epd \theta^{\nu \lambda}$, thus $\tth_0 = \theta$ and $\tth_i = 0$.

We have defined a noncommutative extension of the self-dual model that is (supposedly) dual to the NC-MCS theory.
Duality will be proven next by directly comparing the equations of motion and the algebra of the observables obtained from both models.

The classical equations of motion for the NC-MCS model given by (\ref{ncmcs}) and the NC-SD model disclosed in (\ref{ncsd}) are
\be
	\ep_{\mu \nu \lambda} \partial^\nu \(-\tf^\lambda - \frac 12 \tth^\lambda \tf^\alpha  \tf_\alpha -  \tth^\alpha  \tf_\alpha \tf^\lambda + m A^\lambda \) = 0,
\ee
\be
	f_\mu - \frac 1m \epd \partial^\nu f^\lambda - \frac 12 \tth_\mu f^\alpha  f_\alpha -  \tth^\alpha f_\alpha f_\mu = 0,
\ee
respectively.
Although they look quite distinct at first, the existence of a congruity between these equations of motion follows directly from the correspondence between the basic fields found in (\ref{e10}) which proves that both models describe the same dynamics. Alternatively, by imposing the equality for these equations of motion, the field correspondence given in (\ref{e10}) is reobtained.

To confirm our result, the algebras will be verified. The algebra of the NC-MCS model has been computed in Ref.\cite{ghosh}\footnote{The NC-MCS algebra computed there (see Eq.(14) in \cite{ghosh}) is slightly different from ours due to a minus sign misprint and the absence of the constant $g$ in the initial action.} so we assume it in the sequel. To find the algebra of NC-SD model we shall make use of the symplectic method \cite{symp}. It is immediate to realize that the pre-symplectic matrix for the NC-SD fields $f_\mu$ is singular.  Looking up for the zero-mode, the following constraint is found
\be\label{vinculo}
	- f_0 + \frac 1{m} \ep_{ij} \prt^i f^j + \frac 12 \theta f^i  f_i + \frac 32 \theta f_0 f_0 = 0.
\ee
which is just the zero-component of the field-equations.
In this particular example we may follow two different computational routes with identical consequences: one may either solve the above constraint for the $f_0=f_0(f_i)$ and restrict the computation of the basic brackets for the spatial components $f_i$, whose symplectic matrix is regular or, keep up with the covariant computation by inserting back the constraint (\ref{vinculo}) in the kinetic sector of the Lagrangian and compute the first-iterated symplectic matrix. After that step the (covariant) symplectic matrix becomes nondegenerate, allowing us to extract the generalized brackets (or Dirac brackets) from the entries of its inverse. The result for the basic brackets, from either procedures, is
\ba
	\{ f_0(\vx), f_0(\vy) \}_* & = & g \theta \( f_i(\vx) + f_i(\vy) \) \prt_{(x)}^i \dirac, \nonumber \\
	\{ f_0(\vx), f^i(\vy) \}_* & = & g\( 1 + 3 \theta f_0(\vx) \) \prt_{(x)}^i \dirac + mg\theta \ep^{ij} f_j (\vx) \dirac,  \\
	\{ f^i(\vx), f^j(\vy) \}_* & = & - mg \, \ep^{ij} \dirac. \nonumber
\ea

Following the prescription in eq. (\ref{e10}) we find the algebra for the NC-SD as
\ba
	\{ B(\vx), B(\vy) \} & = & 0, \nonumber \\
	\{ E^i(\vx), B(\vy) \} & = & g\ep^{ij}(1 + \theta B(\vx))\prt_j^{(x)} \dirac, \\
	\{ E^i(\vx), E^j(\vy) \} & = & -gm\ep^{ij}(1 + 2\theta B)\dirac - g\theta ( \ep^{kj} E^i(\vx)+ \ep^{ki} E^j(\vy)) \prt^{(x)}_k \dirac, \nonumber
\ea
where $E^i \equiv - \ep^{ij} \tf_j$ and $B \equiv - \tf_0$.
As expected, it coincides with the algebra for the NC-MCS found in \cite{ghosh}.

\vspace{0.4in}
\section{Conclusion}

In this paper we have studied the issue of duality in the context of the NC-extension of the MCS model up to first-order in the parameter $\theta$.
We have adopted the dual projection approach that has been proved quite useful to study duality in other contexts.
Our basic goal was to find the NC-extension of the Self-Dual model, i.e., the dual companion of the NC-MCS, and to compare our results with the existent studies of the recent literature. Such re-exam of the subject was demanded due to the controversial outputs coming from previous investigations. These studies have approached duality using different techniques. However, none of these studies verified if the candidates to the dual action was able to produce the same set of observable consequences of the original theory. Surprisingly enough, we have not found agreement with any of the previous works.

We have clearly established the dual theory to the noncommutative Maxwell-Chern-Simons theory resulting from the Seiberg-Witten map application to $O(\theta)$. We have found the correspondence among the basic fields and checked that the resulting dual model produces the same set of classical field equations and the same algebra of  observables. This novel dual theory is therefore a natural noncommutative extension of the self-dual model. The duality was proven with direct and transparent procedures and there was no need to resort, for example, to the master Lagrangian approach. Nonetheless, not to prove duality, but to express our result through the traditional approach, we have reinterpret eq. (\ref{e4}) as a master Lagrangian that links aforementioned theories \cite{marcelo} and confirms our results also in this alternative approach. 

An interesting question that remains open is how to relate the duality on commutative fields $A^\mu$ and $f^\mu$ of the NC-MCS and NC-SD theories to a duality between noncommutative fields $\hamu$ and $\hat f^\mu$ \cite{dayi}. If there is an analog map to the Seiberg-Witten map but relating $f^\mu$ with $\hat f^\mu$, it would be possible to extend the present conclusions to the noncommutative fields. Reference \cite{ghosh} suggests this map is $f^\mu = \hat f^\mu$ because, in that paper, ordinary self-dual theory was found as a dual theory to NC-MCS theory (up to first order in $\theta$).
On the other hand, if there were a noncommutative extension of self-dual model with fields $\hat f^\mu$ that is dual to the NC-MCS theory, one could try to find the map between $\hat f^\mu$ and $f^\mu$.
We hope to return to this point in the future.

\vspace{0.3in}
\noindent
{\bf Acknowledgments}
This work is partially supported by PROCAD/CAPES and PRONEX/CNPq.
MSG, JLN and DCR thanks CNPq and FAPERJ (Brazilian research agencies) for financial support. JLN also thanks J. W. Goethe-Uni for financial support.

\end{document}